\documentclass[12pt]{iopart}

\usepackage{cite}
\usepackage[dvips]{graphicx}
\usepackage{color}
\usepackage{ulem}

\newcommand{\bra}[1]{\langle\,{#1}\, |}
\newcommand{\ket}[1]{|\,{#1}\,\rangle}

\begin{document}

\title[Motion of Rydberg-dressed atom clouds]
{Dipole-dipole induced global motion of Rydberg-dressed atom clouds}

\author{M. Genkin, S. W{\"u}ster, S. M{\"o}bius, A. Eisfeld and J.M. Rost}

\address{Max-Planck-Institute for the Physics of Complex Systems, 01187 Dresden, Germany} 
\ead{genkin@pks.mpg.de}

\begin{abstract}
We consider two clouds of ground state alkali atoms in two distinct hyperfine ground states. Each level is far off-resonantly coupled to a Rydberg state, 
which leads to dressed ground states with a weak admixture of the Rydberg state properties. Due to this admixture, for a proper choice of the Rydberg states,
the atoms experience resonant dipole-dipole interactions that induce mechanical forces acting on all atoms within both clouds. 
This behavior is in contrast to the dynamics predicted for bare dipole-dipole interactions between
Rydberg superatoms, where only a single atom per cloud is subject to dipole-dipole induced motion~[Phys. Rev. A {\bf 88} 012716 (2013)].
\end{abstract}
\submitto{\JPB}


\section{Introduction}\label{intro}
The excitation of alkali atoms to Rydberg states is routinely achieved in present day
experiments~\cite{Loew_12,Begu_13,Taus_13,Sass_13,Barr_13,Vite_12,Vinc_13,Pari_12,Trax_13,Dudi_12,Prit_10}. The standard technique is 
a resonant two-photon transition from a ground state $|g\rangle$ to the desired Rydberg state $|r\rangle$ via an intermediate state.
The remarkable properties of Rydberg atoms open a wide area of applications, e.g., as a medium for the implementation of quantum computation protocols~\cite{Saff_10,Isen_10,Molm_11},
a source for single or correlated photons~\cite{Layc_11,Muel_13,Prit_12}, or as quantum simulators for condensed matter systems~\cite{Hagu_12}.
At the same time, the sensitivity of Rydberg atoms to ionization and spontaneous decay may present a problem. Furthermore,
compared to ground state atoms, Rydberg atoms generally require more elaborate trapping techniques~\cite{Ande_11}. To overcome these unfavorable features,
one may apply a continuos off-resonant laser coupling between a ground state and a Rydberg state.
This is referred to as 'Rydberg-dressing'~\cite{Sant_00,Muel_08,Henk_10,John_10,Mayl_10,Li_12,Henk_12,Glae_12,Keat_13}. 
The lifetime of the admixed Rydberg excitation is significantly larger and trapping techniques for ground state atoms may be applicable to a large extent.
On the other hand, Rydberg-dressed ground state atoms will still inherit some of the Rydberg state properties. 
Among these properties, we focus on long-range interactions, in particular resonant dipole-dipole interactions~\cite{Ande_98,Robi_04,Li_05} and van-der-Waals interactions.
While both stem fundamentally from dipole-dipole interactions, 
van-der-Waals interactions arise through off-resonant coupling to nearby quantum states. This approximately results in an interaction term $\hat{V}_{\rm vdW}\sim 1/r^6 \ket{\nu,l;\nu',l'}\bra{\nu,l;\nu',l'}$ 
for a pair of atoms in Rydberg states with given principal quantum number
$\nu$, $\nu'$ and angular quantum numbers $l$, $l'$ separated by a distance $r$. Importantly, this term is diagonal in the electronic state. 
Resonant dipole-dipole interactions, $\hat{V}_{\rm dd}\sim 1/r^3 \ket{\nu,l;\nu',l'}\bra{\nu',l',\nu,l}$ for $|l-l'|=1$, on the other hand, give 
rise to electronic state transfer~\cite{Ates_08,Wust_10,cenap:emergingaggregate}. Both of these interactions can induce atomic motion, but only resonant dipole-dipole interactions link it 
intimately with quantum state transport~\cite{Ates_08,Wust_10}. 
Then the character of motion depends on the overall system eigenstate, 
called exciton, which depends non-trivially on all atom positions.

In the present work, we study the effect of a partial blockade due to van-der-Waals interaction~\cite{Jaks_00,Luki_01,Hern_08} on {\it dressed} resonant dipole-dipole interactions~\cite{Wust_11}. 
Specifically, we consider two atom clouds with radius smaller than the van-der-Waals blockade radius, hence each cloud is
in a full blockade regime. The distance between the clouds is larger than the blockade radius,
and therefore simultaneous Rydberg excitation of one atom from {\it each} cloud is possible.
The Rydberg states are chosen such that
the excited atoms are subject to resonant dipole-dipole interactions, which dominate at these larger distances.
In an earlier article~\cite{moeb_bo_13}, we have considered the same scenario without dressing.  
As demonstrated therein, the dipole-dipole interactions set a single atom pair in motion (one atom from each cloud), so that the initially delocalized Rydberg excitation is in the end localized on the ejected atom. The 
ground state atoms do not move and remain behind.

We show that the dynamics is quite different if the blockaded clouds are weakly Rydberg-dressed rather than excited into a blockade state. Instead of ejecting a single atom, the clouds may move as a whole. 
Whether this occurs and whether motion is attractive or repulsive depends on the systems exciton state, as in the case of bare dipole-dipole interactions.  

In Section~\ref{dressing}, we review the Rydberg dressing scheme and the dipole-dipole interactions between dressed states on a single-atom level~\cite{Wust_11},
by considering a Rydberg dimer with large interatomic separation.
In Section~\ref{setup}, we extend the dressing model to atom clouds and discuss the role of van-der-Waals interactions.
The results of the simulation for the emerging atomic motion
are given in Section~\ref{results}. We summarize our findings and conclude in Section~\ref{concl}.

\section{Binary Dipole-Dipole Interactions and Dressing Scheme}\label{dressing}
Let us first briefly recall bare dipole-dipole interactions between two atoms, both prepared in Rydberg states with principal quantum numbers $\nu,\nu'$ and
angular quantum numbers $l,l'$ satisfying the dipole selection rule $|l-l'|=1$. We assume in the following that the principal quantum number for both atoms
is the same, $\nu=\nu'$ \footnote{For $|\nu-\nu'|>0$, dipole-dipole interaction strengths diminish rapidly.}, and choose $l=0$, $l'=1$.

We abbreviate this two-particle state $\ket{\nu,0;\nu,1}=|sp\rangle$, and the
state with interchanged angular quantum numbers by $|ps\rangle$. 
We further assume that throughout this section the principal quantum number and the interatomic distance ${\bf r}$ are chosen such that van-der-Waals interactions can be neglected.
In this case, the two states couple through a non-vanishing matrix element, giving rise to the dipole-dipole Hamiltonian
\begin{equation}\label{Hdip}
H_{dd}=V_{\rm dd}(r)\left(|sp\rangle\langle ps|+|ps\rangle\langle sp| \right),
\end{equation}
where $V_{\rm dd}(r)=V_0/|{\bf r}|^3$ and $V_0$ is the interaction strength that depends on the transition dipole moment. In general, the interaction is also dependent on the
dipole moment orientation. However, throughout the paper we assume Rydberg states with zero azimuthal quantum number ($m_l=0$) and constrain the atoms in a plane
orthogonal to the quantization axis, which allows us to skip angular dependence. 
One can immediately write down the adiabatic eigenstates of the dipole-dipole Hamiltonian,
\begin{equation}
\varphi^{\pm}=\frac{1}{\sqrt{2}}(|sp\rangle\pm |ps\rangle),
\end{equation}
with the corresponding eigenvalues (adiabatic surfaces) $U^{\pm}({\bf r})=\pm V_{\rm dd}(r)$. The two eigenmodes correspond to attractive and repulsive 
motion of the atoms. This is rather obvious in the adiabatic approximation and can be rigorously demonstrated by solving the time dependent Schr{\"o}dinger equation
for the state $|\Psi({\bf r},t)\rangle=\phi_1({\bf r},t)|sp\rangle+\phi_2({\bf r},t)|ps\rangle$ with the Hamiltonian
\begin{equation}\label{H2atoms}
H=-\sum_{i=1,2}\frac{\hbar^2\nabla_{i}^2}{2M}+H_{dd},
\end{equation}
where $M$ is the atomic mass.

Next, we review the essential features of the dressing scheme. For further details we refer to~\cite{Wust_11}. 
Consider again two alkali atoms, with four essential states $|g\rangle$, $|h\rangle$, $|s\rangle$, and $|p\rangle$. As before, $|s\rangle$ and $|p\rangle$ denote Rydberg
states with angular quantum number 0 and 1 and identical principal quantum number $\nu$, 
while $|g\rangle$ and $|h\rangle$ are two hyperfine ground states in the alkali atom. The Rydberg dressing is achieved
by selectively coupling the states $|g\rangle$ and $|s\rangle$ and, respectively, the states $|h\rangle$ and $|p\rangle$ to each other via far detuned laser fields, as sketched in Figure~\ref{figsetup}.
We denote the effective Rabi frequency of the $|g\rangle$ - $|s\rangle$ transition by $\Omega_s$ and the detuning by $\Delta_s$, and in the same manner we define
$\Omega_p$, $\Delta_p$ for the $|h\rangle$ - $|p\rangle$ transition. 
We will from here on assume the frequencies and detunings to be the same for both transitions, and for simplicity real:
\begin{equation}
\Omega_{s}=\Omega_{p}=\Omega,\quad \Delta_{s}=\Delta_{p}=\Delta.
\end{equation}
With that assumption, the two-atom Hamiltonian encapsulating the laser coupling and the dipole-dipole interaction reads
\begin{equation}\label{Hlas}
H_{\rm dressed}=H_0+H_c,
\end{equation}
where
\begin{equation}
H_0=-\hbar\Delta\sum_{n=1,2}\left(\sigma^n_{ss}+\sigma^n_{pp}\right)+V_{\rm dd}(r)\sum_{n,l=1,2}\left(\sigma^n_{sp}\sigma^l_{ps}\right)
\end{equation}
and
\begin{equation}
H_c=\frac{\hbar\Omega}{2}\sum_{n=1,2}\left(\sigma^n_{gs}+\sigma^n_{sg}+\sigma^n_{hp}+\sigma^n_{ph}\right).
\end{equation}
Here, we have introduced the operators $\sigma_{kk'}^n=|k_n\rangle\langle k'_n|$, where $n$ labels the atom and $k,k'\in\left\{g,h,s,p \right\}$.
As shown in~\cite{Wust_11}, the Hamiltonian~(\ref{Hlas}) can be reduced to an effective one by means of Van Vleck perturbation theory,
which ultimately allows to introduce dressed dipole-dipole interactions between
dressed ground states $|\tilde{g}\rangle$, $|\tilde{h}\rangle$,
\begin{equation}
|\tilde{g}\rangle={\cal N}(|g\rangle+\alpha\,|s\rangle),\quad|\tilde{h}\rangle={\cal N}(|h\rangle+\alpha\,|p\rangle),
\end{equation}
where $\alpha=\Omega/(2\Delta)\ll 1$ is a dimensionless scaling parameter and ${\cal N}=1/\sqrt{1+\alpha^2}$ a normalization factor.
We define two-particle states $|\tilde{g}\tilde{h}\rangle$ and $|\tilde{h}\tilde{g}\rangle$, corresponding to the first atom being
in the state $|\tilde{g}\rangle$ and the second atom in $|\tilde{h}\rangle$, and vice versa. 
In this basis, the effective Hamiltonian takes the form
\begin{equation}\label{Hdressed}
\tilde{H}_{dd}=\tilde{V}\left(|\tilde{g}\tilde{h}\rangle\langle \tilde{h}\tilde{g}|+|\tilde{h}\tilde{g}\rangle\langle \tilde{g}\tilde{h}| \right)
+\hbar\tilde{W}\left(|\tilde{g}\tilde{h}\rangle\langle \tilde{g}\tilde{h}|+|\tilde{h}\tilde{g}\rangle\langle \tilde{h}\tilde{g}| \right)
\end{equation}
with
\begin{equation}\label{dresseddip}
\tilde{V}=V_{\rm dd}(r)\left[\frac{1}{1-\left(\frac{V_{\rm dd}(r)}{2\hbar\Delta}\right)^2}\right]
\end{equation}
and
\begin{equation}\label{offdiag}
\tilde{W}=2\alpha^2\Delta+2\alpha^4\Delta\left[\frac{1}{1-\left(\frac{V_{\rm dd}(r)}{2\hbar\Delta}\right)^2}-2\right]. 
\end{equation}
We consider interatomic distances $|{\bf r}|\gg r_c=(V_0/2\hbar\Delta)^{1/3}$, and hence we can neglect 
the position dependence of $\tilde{W}$, i.e., in
the diagonal terms of $\tilde{H}_{dd}$. The remaining diagonal terms (light shift) depend solely on the
dressing laser parameters and have no further effect on exciton transport or inter-atomic forces. Hence, considering only the first (off-diagonal) term in Eq.~(\ref{Hdressed}), we see that the effective Hamiltonian 
for the dressed two-atom system has exactly the same structure as the one in Eq.~(\ref{Hdip}), with a
dressing-dependent scaling of the interaction strength.

Illustratively, one can think of Eq.~(\ref{dresseddip}) as arising from a three-step-process: In the first step, the ground states $|g\rangle$, $|h\rangle$ are excited to the
Rydberg states $|s\rangle$, $|p\rangle$ by the dressing laser. In the second step, the dipole-dipole interactions flip the Rydberg states $|s\rangle$ and $|p\rangle$.
In the final step, the Rydberg states are de-excited back to the respective ground states by the dressing field (see Fig.~\ref{figsetup}).
This involves four photoinduced transitions in total, hence the interaction strength scales as $\alpha^4$.
We can thus conclude that the dipole-dipole induced motional dynamics for a single pair of Rydberg-dressed atoms at large distances is indeed analogous to the dynamics
for genuine Rydberg atoms, the dressing merely leads to a rescaling of the Hamiltonian matrix elements and hence to different time scales of motion.

\section{Dressed atom clouds}\label{setup}
\begin{figure}
\begin{footnotesize}
\begin{center}
\scalebox{0.99}{\includegraphics{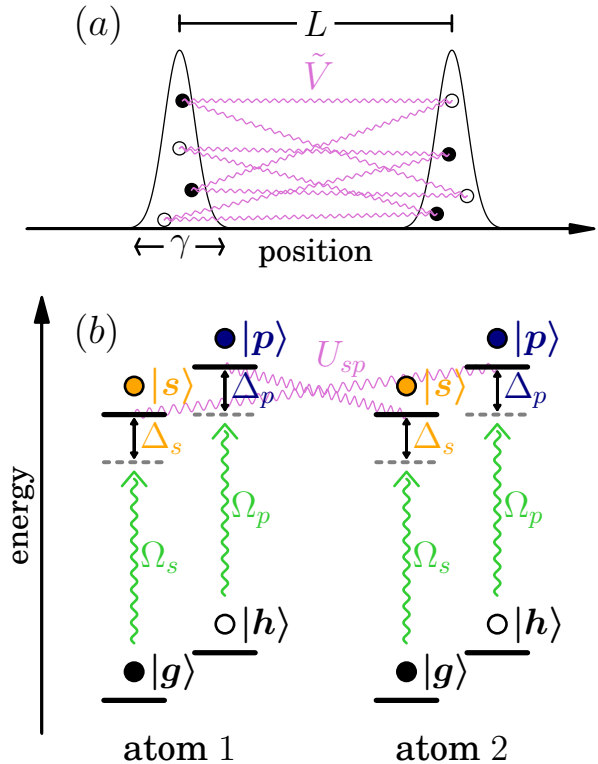}}\\
\end{center}
\begin{quote}
\caption{(Color online) Schematic of the setup.
Lower panel: Level-scheme for the dressed dipole-dipole interactions between two single atoms. Upper panel: The Rydberg-dressed atoms within one cloud
interact with all atoms from the other cloud for which the Rydberg states satisfy the dipole selection rule.
}
\label{figsetup}  
\end{quote}
\end{footnotesize}
\end{figure}

Having discussed dressed dipole-dipole interactions for a single atom pair, we now consider a more complicated situation
where the two atoms are replaced by Rydberg-dressed atom clouds. The spatial extension $\gamma$ of these clouds is taken to be
smaller than the van-der-Waals blockade radius, while the distance $L$ between the two clouds is larger (see Fig.~\ref{figsetup}a). As will be elaborated below,
in this way we suppress dressed dipole-dipole interactions within each cloud, but allow these interactions for any two atoms
from different clouds.

For simplicity, we take the total number of atoms $N$ to be even, with $N/2$ atoms in each cloud.
The two clouds are labeled by A and B, and we define index sets $\mathcal{A}, \mathcal{B}$
for atoms residing in cloud A or B, respectively.
The Hamiltonian is a straight-forward extension of the one of section~\ref{dressing}, Eq.~(\ref{Hlas}).
In this scenario, van-der-Waals interactions can no longer be neglected since the interatomic distances within each cloud are small.
Hence, the full electronic many-body Hamiltonian includes the laser dressing as well as both the dipole-dipole interactions and van-der-Waals interactions.
Explicitly, it reads
\begin{equation}\label{helec}
H_{\rm el}=H_{\rm las}+H_{\rm int},
\end{equation}
where
\begin{equation}
H_{\rm las}=\hbar\sum_{n=1}^{N}\left[\frac{\Omega}{2}\left(\sigma^n_{gs}+\sigma^n_{sg}+\sigma^n_{hp}+\sigma^n_{ph}\right)-\Delta\left(\sigma^n_{ss}+\sigma^n_{pp}\right)\right]
\end{equation}
and
\begin{equation}\label{hint}
H_{\rm int}=\sum_{n,l=1}^{N}\left[\frac{V_0}{|{\bf r}_n-{\bf r}_l|^3}\sigma^n_{sp}\sigma^l_{ps}+\sum_{a,b=s,p}\frac{C^{ab}_6}{|{\bf r}_n-{\bf r}_l|^6}\sigma^n_{ab}\sigma^l_{ab}\right].
\end{equation}
In the last equation, $C^{ab}_6$ denotes the state-dependent van-der-Waals interaction strength.
Here, $H_{\rm las}$ contains all terms arising from the external fields, while $H_{\rm int}$ accounts for atomic interactions. In the Hamiltonian for just two atoms, Eq.~(\ref{Hlas}),
we chose a slightly different grouping of terms in order to facilitate the perturbative treatment, but we emphasize once more that the two Hamiltonians only differ in so far as the many-body Hamiltonian~(\ref{helec})
also includes van-der-Waals interactions, which were neglected in Eq.~(\ref{Hlas}) due to the assumption of large distances. 

Let us first consider the effect of the van-der-Waals interactions. It is well known that at small interatomic distances, 
these interactions lead to an energy offset of all many-particle states with more than one Rydberg excitation with respect to the energy of the states with just a single excitation.
Therefore, an external laser field tuned resonantly to the atomic transition from a ground state to a Rydberg state can create at most one Rydberg excitation within 
a radius at which the van-der-Waals energy offset is larger than the laser linewidth. The linewidth is determined by the Rabi frequency of the transition, which provides
an estimate for the blockade radius $r_{\rm bl}\approx (C_6/\hbar\Omega)^{1/6}$. This effect gives rise to so called superatomic states~\cite{Dudi_12, Heid_07, Robi_05}  where several atoms confined
in a volume $\sim 4\pi r^3_{\rm bl}/3$ coherently share a single Rydberg excitation. In an earlier paper, we have studied motional dynamics induced by resonant dipole-dipole interactions
between such superatoms~\cite{moeb_bo_13} without dressing. Interestingly, the delocalized coherently shared excitation does not prevail - the dipole-dipole forces eject a single atom from each superatom,
localizing the entire Rydberg excitation on this atom. For Rydberg-dressed atoms, the concept of van-der-Waals blockade needs to be slightly
modified, as the dressing blockade radius is now defined as the interatomic distance at which the dressing field is rendered ineffective by the van-der-Waals interaction.
Since the dressing lasers are already far detuned ($\Omega\ll\Delta$), the dressing blockade radius can be estimated as $\tilde{r}_{\rm bl}\approx(C_6/2\hbar|\Delta|)^{1/6}$~\cite{Henk_10}.

We recall that in our setup the spatial extension of each cloud is smaller than the blockade radius by construction. Consequently, each cloud can only sustain a single Rydberg excitation.
This implies that (dressed) dipole-dipole interactions between an atom pair $nm$ with $n,m\in\mathcal{A}$ (or $n,m\in\mathcal{B}$ respectively) are suppressed as this would require
two excitations within the blockade radius, while for an atom pair with $n\in\mathcal{A},m\in\mathcal{B}$ (or vice versa) the dipole-dipole interactions remain possible. 

The model~(\ref{hint}) is oversimplified
at very small atomic distances, where the interaction potentials between adjacent Rydberg levels approach each other very closely in energy,
displaying multiple avoided crossings (see e.g. Figure 1 in~\cite{Schwett_06}). Nonetheless it captures the only essential physics, which is that the blockade also holds
for very small interatomic separations, as confirmed by experiment. Here, due to the steepness of the molecular potentials excitation to any Rydberg pair state (ss,sp,pp) is strongly
suppressed~\cite{Ates_12}.

This is taken into account by a reduction of the many-body Hilbert space, removing all states that contain more than one Rydberg atom in the same cloud.
For the Hamiltonian~(\ref{helec}) with $C_6^{(a,b)}=0$, we then apply van-Vleck perturbation theory~\cite{Vleck_29,Shav_80,Wust_11} 
to derive an effective Hamiltonian in the {\it ground-state manifold}, 
the many-body space spanned by $|\tilde{g}\rangle$, $|\tilde{h}\rangle$ for each atom. We construct this many-body basis perturbatively in the dressing parameter $\alpha$, which is also the small parameter of 
van-Vleck perturbation theory. Hence we write for example $|\tilde{g}\tilde{h}\rangle \sim |gh\rangle + \alpha |gp\rangle + \alpha  |sh\rangle $, {\it without} the presence 
of the possibly blockade forbidden contribution $\alpha^2 |sp\rangle $. With methods as used in~\cite{Wust_11} we then obtain the dressed electronic Hamiltonian
\begin{equation}\label{heltilde}
\tilde{H}_{\rm el}=\sum_{n\in \mathcal{A}, m\in \mathcal{B}}\tilde{V}_{nm}(r_{nm})[\sigma_{\tilde{g}\tilde{h}}^n\sigma_{\tilde{h}\tilde{g}}^m+\sigma_{\tilde{g}\tilde{h}}^m\sigma_{\tilde{h}\tilde{g}}^n],
\end{equation}
where $r_{nm}$ denotes the interatomic distance $|{\bf r}_n-{\bf r}_m|$ and $\tilde{V}_{nm}(r_{nm})$ is the dressed two-body dipole-dipole interaction
from Eq.~(\ref{dresseddip}). Since the distance of the clouds obeys $L\gg r_c$, 
we may even approximate \begin{equation}
\label{dresseddipapprox}
\tilde{V}_{nm}(r_{nm})\approx\frac{V_0\alpha^4}{r_{nm}^3}.
\end{equation}
From the Hilbert space in which the Hamiltonian of Eq.~(\ref{heltilde}) operates, all doubly excited states within one cloud have been removed.
Consequently, our approach will only be self-consistent if $\alpha^2 N\ll 1$, since the dressing-induced excited state occupation per cloud is roughly given by $\alpha^2 N$.

\section{Atomic motion}\label{results}
Having defined the state space and the Hamiltonian for the electronic degrees of freedom, we are ready to tackle the full dynamics of the system, including atomic motion.
To this end, we consider the many-body version of the Hamiltonian given in Eq.~(\ref{H2atoms}) for dressed interactions, i.e.,
\begin{equation}\label{fullHamiltonian}
\tilde{H}=-\sum_{i=1}^N\frac{\hbar^2\nabla_{i}^2}{2M}+\tilde{H}_{\rm el},
\end{equation}
with $\tilde{H}_{\rm el}$ from Eq.~(\ref{heltilde}).
As an illustration, let us consider a simple case of four atoms in total (two in each cloud), and restrict the position space to one dimension for each atom. 
We consider larger atom numbers in the Appendix. 
We assume that initially two of the atoms are in the dressed state $|\tilde{g}\rangle$, and the other two atoms in $|\tilde{h}\rangle$. 
This choice is made for a better illustration, in particular since the dressed dipole-dipole
interactions conserve the total number of atoms in either of these states. Proposals regarding the possibility to dynamically create
an initial state with a given distribution of atoms on the states $|\tilde{g}\rangle$ and $|\tilde{h}\rangle$
are given in Ref.~\cite{Moeb_13}.
With our choice, the electronic Hilbert space
is spanned by the states $\{|\tilde{g}\tilde{g}:\tilde{h}\tilde{h}\rangle,|\tilde{g}\tilde{h}:\tilde{g}\tilde{h}\rangle,
|\tilde{g}\tilde{h}:\tilde{h}\tilde{g}\rangle,|\tilde{h}\tilde{g}:\tilde{g}\tilde{h}\rangle,|\tilde{h}\tilde{g}:\tilde{h}\tilde{g}\rangle,|\tilde{h}\tilde{h}:\tilde{g}\tilde{g}\rangle\}$.
In this notation, the colon separates the one-particle states within cloud A and B, respectively. As explained above, tensor-products are defined only up to order $\alpha$.
For a better visualization, we also give the matrix elements of $\tilde{H}_{\rm el}$ in this basis,
\begin{equation}\label{Hmatrixform}
\tilde{H}_{\rm el}=\left(\begin{array}{cccccc}
0 & \tilde{V}_{23} & \tilde{V}_{24} & \tilde{V}_{13}  & \tilde{V}_{14}  & 0 \\
\tilde{V}_{23} & 0 & 0 & 0 & 0 & \tilde{V}_{14}  \\
\tilde{V}_{24} & 0 & 0 & 0 & 0 & \tilde{V}_{13}  \\
\tilde{V}_{13} & 0 & 0 & 0 & 0 & \tilde{V}_{24}  \\
\tilde{V}_{14} & 0 & 0 & 0 & 0 & \tilde{V}_{23}  \\
0 & \tilde{V}_{14} & \tilde{V}_{13} & \tilde{V}_{24}  & \tilde{V}_{23}  & 0
\end{array}\right).
\end{equation}
Already at this point, we note that the structure and sparsity of this Hamiltonian differ from the case of a genuine coherently shared Rydberg excitation, cf. Eq.~(6) and Table I. in Ref.~\cite{moeb_bo_13},
and hence we may indeed expect a different kind of motional dynamics for dressed ground state atoms.

In order to determine the expected motional dynamics, we inspect the Born-Oppenheimer surfaces of the system, i.e., the eigenvalues of the electronic Hamiltonian~(\ref{heltilde}).
They are given by
\numparts
\begin{eqnarray}
U_1({\bf R})&=&0, \\ 
U_2({\bf R})&=&0, \\
U_3({\bf R})&=& \sqrt{(\tilde{V}_{14}-\tilde{V}_{23})^2+(\tilde{V}_{13}-\tilde{V}_{24})^2}, \\
U_4({\bf R})&=& -\sqrt{(\tilde{V}_{14}-\tilde{V}_{23})^2+(\tilde{V}_{13}-\tilde{V}_{24})^2}, \\
\label{u5}
U_5({\bf R})&=& \sqrt{(\tilde{V}_{14}+\tilde{V}_{23})^2+(\tilde{V}_{13}+\tilde{V}_{24})^2}, \\
\label{u6}
U_6({\bf R})&=& -\sqrt{(\tilde{V}_{14}+\tilde{V}_{23})^2+(\tilde{V}_{13}+\tilde{V}_{24})^2},
\end{eqnarray}
\endnumparts
where ${\bf R}$ is a vector containing all atomic positions.
In a regime where the atomic motion is adiabatic, the gradient of the k$th$ eigenvalue with respect to the position $r_i$ of the i$th$ atom determines the force and hence 
the motion of this atom~\cite{Ates_08} on the k$th$ surface,
\begin{equation}\label{forces}   
F_i^k=-\partial_{r_i}U_k({\bf R}).
\end{equation}
Let us for clarity's sake consider the special case where the distances between all atom pairs with atoms in different clouds are the same,
i.e., $r_{13}=r_{14}=r_{23}=r_{24}\equiv \varrho$. Then the gradient of the eigenvalues $U_{5}$ and $U_{6}$ has the structure
\begin{equation}\label{gradsurf}
{\bf F}^5=\left(\begin{array}{cccc}
\xi \\ \xi \\ -\xi \\ -\xi\end{array}\right)=-{\bf F}^6,
\end{equation}  
with the force $\xi=3\sqrt{2} V_0\alpha^4/\varrho^4$ and $V_0$ as defined below Eq.~(\ref{Hdip}).
\begin{figure}
\begin{footnotesize}
\begin{center}
\scalebox{0.35}{\includegraphics{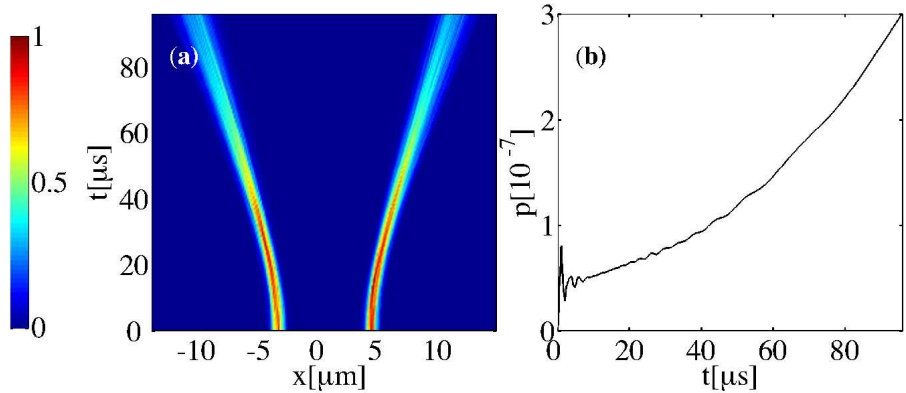}}\\
\end{center} 
\begin{quote}
\caption{(Color online)
Motion of dressed ground state atoms, prepared in a repulsive configuration. (a): Total atom density as a function of time.
Clouds of Rydberg-dressed ground state atoms are set in motion. (b): Non-adiabatic population transfer from the adiabatic
surface on which the dynamics is initiated is of the order of $10^{-7}$.
The dynamics is shown for 8 atoms in total.
The following parameters are used: $M=12800\,$a.u., $V_0=1.08\cdot 10^6\,$a.u., $\gamma=0.5\,\mu$m, $L=8\,\mu$m, $\Omega=4\,$MHz, $\alpha=0.18$.
The parameters for the mass and interaction strength correspond to $\nu=36$ states in $^7$Lithium, the interaction strength is obtained from the scaling $V_0\approx\mu_0^2\nu^4$
with $\mu_0\approx 0.8\,$a.u. The lifetime of the dressed Rydberg state with $\alpha=0.18$ is $\tau\approx 1\,$ms, calculated as
$\tau\approx \tau_0\nu^3/\alpha^2$ with $\tau_0\approx 3\cdot 10^7\,$a.u.~\cite{book:gallagher,Theo_84}. The lifetime is therefore an order of magnitude
larger than the timescale on which the motional dynamics takes place. 
}
\label{fig2}
\end{quote}
\end{footnotesize}
\end{figure}
We see that the two atoms in cloud $A$ experience the same force as the two atoms in cloud $B$ but with opposite sign.
In other words, the dynamics on these two adiabatic surfaces corresponds to repulsion and attraction of the {\it whole} cloud. In the general case the interatomic distances are
not exactly the same. Hence the matrix elements $\tilde{V}_{ij}$ will slightly differ, and so will the forces. However, the qualitative 
picture remains the same as long as the spatial extension of each cloud is much smaller than the distance between the clouds. The treatment can also be extended to larger atom numbers, see Appendix.
Here, we have numerically simulated the atomic motion for $N=8$. 

While a full quantum mechanical solution of the time-dependent Schr{\"o}dinger equation with the Hamiltonian from Eq.~(\ref{fullHamiltonian}) is not feasible for a large number of atoms,
quantum-classical hybrid methods can often be successfully applied to systems such as considered here. 
Among those, Tully's surface hopping algorithm is a well-established approach~\cite{Tull_71,Tull_90,Hamm_94,Ates_08,Wust_10,Moeb_11,moeb_bo_13}, which
also allows to estimate the relevance of non-adiabatic effects. 
As the method is described in the aforementioned references, here we will only briefly mention a few most crucial aspects.
In the framework of Tully's algorithm, the electronic degrees of freedom are treated quantum mechanically, while the motion of the atoms is treated classically. All physical quantities are
derived by averaging over a large amount of such trajectories, each of which is propagated on a single adiabatic surface. The presence of non-adiabatic effects is incorporated into the dynamics
by means of stochastic switches between different adiabatic surfaces in each timestep. 
The initial conditions for the propagation are chosen such that they resemble the Wigner function of the initial quantum state.

The results of the simulation, which constitute the main finding of the present work, are shown in Figure~\ref{fig2}.
We observe that the dressed dipole-dipole interactions set all atoms in motion (left panel). This is in contrast to the dynamics found for atom clouds coherently sharing a single genuine
Rydberg excitation~\cite{moeb_bo_13}, where ultimately only a single atom pair is ejected. We also find that the dynamics is highly adiabatic, as the 
non-adiabatic population transfer is of the order of $10^{-7}$ (right panel). 
The reason for the entirely different behavior between coherently excited, blockaded clouds and weakly Rydberg-dressed clouds can
be understood by inspecting the corresponding Hamiltonian of the electronic degrees of freedom.
A coherently shared excitation allows dipole-dipole interactions between {\it one single atom pair at a time}. 
In the dressed case, in contrast, {\it all} (dressed) ground state
atoms within one cloud interact with {\it all} atoms from the other cloud.

We also observe a spread of the initial wave packet from the time evolution of the spatial density in Figure~\ref{fig2}. In order to assess its origin,
we first estimate the intrinsic quantum mechanical dispersion for a single atom. The time-dependent width $\gamma(t)$ is given by
$\gamma(t)=\gamma_0\sqrt{1+\beta(t)^2}$, where $\gamma_0$ is the initial width and $\beta(t)=\hbar t/(2M\gamma_0^2)$. Inserting the parameters
of our simulation ($M=12800\,$a.u., $\gamma_0=0.5\,\mu$m) and an evolution time of $T=100\,\mu$s, we obtain
$\gamma(T)\approx 0.84\,\mu$m. This gives a visible contribution to the wave packet spreading, however, it cannot fully explain the total final width.
An additional factor is the dipole-dipole interaction between the wave packets. Its effect on the spreading is simply explained in the picture of single trajectories.
Let us consider two pairs of trajectories, one pair initially starting in the inner tail of each wave packet (with respect to the position of the other one) and another
pair in the outer tail. Due to the $1/r^3$ dependence of the dipole-dipole interactions, the former two trajectories experience a stronger force. The interaction is,
in our case, repulsive, which means that they acquire a higher velocity and will eventually overtake the trajectories initiated in the outer tail. This leads
to a temporary asymetric squeezing of the wavepacket, but eventually the inhomogenious velocity distribution becomes an additional factor which increases the dispersion.
For the parameters considered in our simulation, this effect and the intrinsic quantum mechanical spreading are of similar importance.

Note that the simple mechanical repulsion of the clouds of atoms can also be achieved via dressed van-der-Waals interactions~\cite{Henk_10,Pupi_10}, when
the radius of each cloud is less than the blockade radius $\tilde{r}_{bl}$, but their separation larger.
Dressed dipole-dipole-interactions discussed here are more complicated, but offer features that cannot be realized with van-der-Waals dressing:
We can switch between attraction, Eq.~(\ref{u6}) and repulsion, Eq.~(\ref{u5}), by initializing a different electronic state. For van-der-Waals dressing this would require addressing
different Rydberg states~\cite{Mauc_11}, and hence changing the dressing setup. Due to the electronic state dependence of the motion for dressed dipole-dipole interactions, 
they can be employed to study linked exciton and motional dynamics~\cite{Wust_10,Moeb_11,Leon_13} and even create mesoscopic entangled states~\cite{Moeb_13}.

Here we have exclusively focused on unconfined interacting atoms. Dressed dipole-dipole interactions between atoms confined in an optical lattice also offer intriguing opportunities, 
such as engineering exciton-phonon interaction Hamiltonians~\cite{Hague_12}. 
In such as case, the importance of non-adiabatic decoherence effects has to be assessed carefully~\cite{Li_13,Macr_14}.

\section{Summary and Conclusion}\label{concl}
We have studied the effects of the Rydberg blockade on the motion of Rydberg-dressed atom clouds, induced by dressed dipole-dipole interactions. 
We predict a global motion for all atoms, explicitly demonstrated numerically using Tully's quantum-classical hybrid method.
The result can be qualitatively understood by examining the corresponding adiabatic surfaces. 
Such a dynamics stands in contrast to
the one of atom clouds coherently sharing a Rydberg excitation, where only a single pair of atoms is set in motion.
The observed behavior paves the way towards the realization of
entangled mesoscopic motional states, where the entanglement can prevail for microseconds over distances of several micrometers~\cite{Moeb_13}.
A possible implementation of momentum and entanglement transport in one-dimensional Rydberg chains~\cite{Wust_10,Moeb_11}, but with dressed atom clouds as sites as we consiedered here,
would emphasize differences between pure van-der-Waals dynamics and the one stemming from exciton transport~\cite{cenap:emergingaggregate}. 

\ack
We are happy to thank Klaus M{\o}lmer, Pierre Pillet, Nils Henkel and Thomas Pohl for helpful discussions.
Financial support by the Marie Curie Initial Training Network 'COHERENCE' is gratefully acknowledged.

\appendix
\section{Extension to large atom numbers}
In this appendix, we discuss the setup sketched in Figure~\ref{figsetup} for larger numbers of atoms than treated in the man text.
In particular, we consider the electronic basis, the adiabatic surfaces,
as well as the scaling of the force with atom number for two dressed atom clouds.

We have $N$ atoms in total, each of which can be in either $|\tilde{g}\rangle$ or $|\tilde{h}\rangle$, which leads to $2^N$ possible many-body states. 
However, a choice of a given initial state immediately reduces the dimension substantially, since the dressed dipole-dipole interactions conserve
the total number of atoms in either of the states. For example,
if the initial state is prepared such
that $N/2$ atoms are in either of these two states, the dimension of the Hilbert space reduces to ${\cal D}=\left(\begin{array}{c}N \\ N/2 \end{array}\right)$.
For convenience, we  take the number of atoms in the two clouds to be the same and define $n\equiv N/2$~\footnote{The essential features of the dynamics, such as the existense of adiabatic surfaces
leading to global attractive/repulsive motion of the atom clouds, also prevail if the number of atoms within each cloud is not the same.}. 
The many-body states can be classified further in terms of the number of atoms in a given state and cloud. We denote the number
of atoms in cloud A, which are in the state $|\tilde{h}\rangle$ by $N_h$. For a given $N_h$, there exist $\left(\begin{array}{c}n \\ N_h \end{array}\right)^2$ states. Note that the equality
${\displaystyle \sum_{N_h}} \left(\begin{array}{c}n \\ N_h \end{array}\right)^2=\left(\begin{array}{c}N \\ n \end{array}\right)$ holds.
Next, we consider the matrix structure of the many-body Hamiltonian (Eq.~(\ref{heltilde})), using the basis states classified according to $N_h$. 
One can write the Hamiltonian in terms of blocks $B_{ij}$, where each block is a matrix of dimension
\begin{equation}
{\rm dim} B_{ij}=\left(\begin{array}{c}n \\ i \end{array}\right)^2\,\times\,\left(\begin{array}{c}n \\ j \end{array}\right)^2,
\end{equation}
formally coupling basis states from different $N_h$-manifolds with $N_h=i$ and $N_h=j$,
\begin{equation}
\tilde{H}_{\rm el} =  \left(\begin{array}{cccc}B_{0,0} & B_{0,1} & \cdots & B_{0,n} \\
B_{1,0} & B_{1,1} & \cdots & B_{1,n} \\
\vdots & \vdots & \ddots & \vdots \\
B_{n,0} & B_{n,1} & \cdots & B_{n,n}
\end{array}\right).
\end{equation}
However, the dressed dipole-dipole interactions can only flip
an atom pair from $|\tilde{g}\tilde{h}\rangle$ to $|\tilde{h}\tilde{g}\rangle$, as long as the two involved atoms reside in different clouds, due to the blockade condition.
This means that the matrix elements of the Hamiltonian can only be non-zero between state manifolds for which $N_h$ differs exactly by one. Hence, the only non-vanishing blocks
are adjacent to the main diagonal, and the Hamiltonian assumes the form
\begin{equation}
\tilde{H}_{\rm el} = \left(\begin{array}{ccccc}0 & B_{0,1} & \cdots & 0 & 0 \\
B_{1,0} & 0 & \cdots & 0 & 0 \\ 
\vdots & \vdots & \ddots & \vdots & \vdots \\  
0 & 0 & \cdots & 0 & B_{n-1,n} \\
0 & 0 & \cdots & B_{n,n-1} & 0
\end{array}\right), 
\end{equation}
with $B_{ij} = B_{ji}^\dagger$. One can further give the sparsity of non-vanishing blocks. For a state within a given $N_h$-manifold, there are $\left(\begin{array}{c}n \\ N_h \end{array}\right)^2(n-N_h)^2$
non-zero matrix elements leading to a state within the $N_h+1$-manifold and $\left(\begin{array}{c}n \\ N_h \end{array}\right)^2N_h^2$ non-zero matrix elements leading to a state within the $N_h-1$-manifold.
Each element has the form as given in Eqs.~(\ref{dresseddip}), (\ref{dresseddipapprox}).

Having established the structure of the Hamiltonian, we proceed with a closer examination of the adiabatic surfaces. We consider the eigenvalue equation
\begin{equation}\label{EVs}
\tilde{H}_{\rm el}|\psi_m\rangle=U_m|\psi_m\rangle,
\end{equation}
and expand each eigenstate $|\psi_m\rangle$ in terms of the basis states classified by $N_h$,
\begin{equation}
|\psi_m\rangle=\sum_{N_h=0}^n{\bf c}^m_{N_h}|\{\varphi_{N_h}\}\rangle.
\end{equation}
Here, $|\{\varphi_{N_h}\}\rangle$ symbolically denotes all basis states with a given $N_h$ and ${\bf c}^m_{N_h}$ is the coefficient vector which gives the contribution of each single basis state to the $m$-th eigenstate.
For simplicity, throughout the appendix we assume that all atoms within one cloud are at the same position, implying that all non-zero matrix elements of $\tilde{H}_{\rm el}$ have the same magnitude, which we abbreviate by
$W=W({\bf R})\approx V_0\alpha^4/L^3$ in the following. 
With this assumption, one finds eigenstates where all entries in the ${\bf c}^m_{N_h}$ have the same magnitude (denoted by $c^m_{N_h}$), which allows us to rewrite the eigenvalue 
equation~(\ref{EVs}) as
\begin{equation}
U_mc^m_{N_h}  = W(N_h^2 c_{N_h-1}^m + (n-N_h)^2 c_{N_h+1}^m).
\end{equation}
With the additional rescaling 
\begin{equation}
\tilde{c}^{\,m}_{N_h}= \left(\begin{array}{c}n \\ N_h \end{array}\right)c^m_{N_h},
\end{equation}
the new coefficients are normalized,
\begin{equation}
\sum_{N_h} |\tilde{c}^{\,m}_{N_h}|^2=1,
\end{equation}
and we arrive at
\begin{equation}
\tilde{U}_m \tilde{c}^{\,m}_{N_h} = (A_{N_h} \tilde{c}^{\,m}_{N_h-1} + B_{N_h} \tilde{c}^{\,m}_{N_h+1}),
\end{equation}
where
\begin{eqnarray}
\tilde{U}_m &=& \frac{U_m}{W}, \\
A_{N_h} &=&  N_h(n-N_h+1), \\
B_{N_h} &=& N_h(N_h+1). 
\end{eqnarray}
We can also express the re-scaled adiabatic surfaces as
\begin{equation}\label{surfacerescaled}
\tilde{U}_m=\sum_{N_h=0}^{n-1}2B_{N_h}{\rm Re}\left\{\left(\tilde{c}^{\,m}_{N_h}\right)^{*}\tilde{c}^{\,m}_{N_h+1}\right\}.
\end{equation}
The adopted transformations effectively reduce the dimension of the problem from $\left(\begin{array}{c}N \\ n \end{array}\right)$  to $n+1$ essential states,
which is now easily accessible numerically. For example, we can estimate the scaling of the adiabatic surfaces corresponding to an attraction/repulsion of all atoms with increasing atom number as
$U_{\rm rep/att}\approx n^2W/2$.

Finally, we consider the force acting on a single atom, according to Eq.~(\ref{forces}). We can rewrite the force acting
on the $i$-th atom on the $m$-th surface as
\begin{equation}\label{HellFeyn}
{\bf F}^m_i=-\langle\psi_m|\partial_{{\bf r}_i}\tilde{H}_{\rm el}({\bf R})|\psi_m\rangle,
\end{equation}
where $|\psi_m\rangle$ is the eigenstate corresponding to the $m$-th surface. Since by assumption the entries of $\tilde{H}_{\rm el}$ have the same magnitude,
we only have to count the number of matrix elements which depend on ${\bf r}_i$ in order to evaluate Eq.~(\ref{HellFeyn}). After some algebraic transformations, the explicit expression becomes
\begin{equation}
-{\bf F}^m_i=\partial_{{\bf r}_i}W({\bf R})\frac{1}{n} \sum_{N_h=0}^{n-1}2B_{N_h}{\rm Re}\left\{\left(\tilde{c}^{\,m}_{N_h}\right)^{*}\tilde{c}^{\,m}_{N_h+1}\right\}.		
\end{equation}    
By comparing the expression with Eq.~(\ref{surfacerescaled}), we arrive at
\begin{equation}
-{\bf F}^m_i=\partial_{{\bf r}_i}W({\bf R})\frac{\tilde{U}_m}{n}.
\end{equation}
Since we previously found the scaling $U_{\rm rep/att}\sim n^2$ for attractive/repulsive surfaces, from the above equation we infer that the force per atom scales linearly with the number of atoms.

\section*{References}
\bibliographystyle{unsrt.bst}
\bibliography{dressed_v7.bib}
\end{document}